\documentclass[prl,twocolumn,superscriptaddress]{revtex4}

\usepackage{amsfonts,amsmath,amssymb,amsthm}
\usepackage{epic,eepic,graphicx,mathpple}

\usepackage[usenames,dvipsnames]{color}

\usepackage{amsmath, amsthm, amssymb}
\usepackage{graphicx}

\theoremstyle{definition}

\theoremstyle{remark}

\newcommand{\eeee}{\textrm{ee}}
\newcommand{\eeii}{\textrm{ei}}
\newcommand{\iiii}{\textrm{ii}}
\newcommand{\iiee}{\textrm{ie}}
\newcommand{\ee}{\textrm{e}}
\newcommand{\ii}{\textrm{i}}

\newcommand{\vvec}[1]{{\bf #1}}

\newcommand{\newstuff}[1]{{#1}}

\begin{document}


\title{Spatiotemporal dynamics and reliable computations in recurrent spiking neural networks}

\author{Ryan Pyle}
\affiliation{Department of Applied and Computational Mathematics and Statistics, University of Notre Dame, Notre Dame, IN 46556}
\author{Robert Rosenbaum} 
\affiliation{Department of Applied and Computational Mathematics and Statistics, University of Notre Dame, Notre Dame, IN 46556}
\affiliation{Interdisciplinary Center for Network Science and Applications, University of Notre Dame, Notre Dame, IN 46556}

\begin{abstract}
Randomly connected networks of excitatory and inhibitory spiking neurons provide a parsimonious model of neural variability, but are notoriously unreliable for performing computations.  We show that this difficulty is overcome by incorporating the well-documented dependence of  connection probability on distance.  Spatially extended spiking networks exhibit symmetry-breaking bifurcations and generate  spatiotemporal patterns that can be trained to perform dynamical computations.  

\end{abstract}

\maketitle

Biological neuronal networks exhibit irregular and asynchronous activity~\cite{Softky1993,Shadlen1994} {that is often modeled using randomly connected networks of  excitatory and inhibitory spiking neurons.  In these models, an approximate balance between excitation and inhibition creates a push-pull dynamic that combines with random connectivity and chaotic network dynamics to produce asynchronous-irregular spiking activity similar to that observed in experimental recordings~\cite{vanVreeswijk:1996us,vanVreeswijk:1998uz,Brunel:2000th,Renart2010}.}

Despite their ability to explain the genesis of neural variability, asynchronous-irregular spiking network models have a critical shortcoming: Their microscopic dynamics -- at the level of spike times -- are intricate and nonlinear, but largely unreliable~\cite{vanVreeswijk:1996us,vanVreeswijk:1998uz,Monteforte2012,Lajoie2013}.  Their macroscopic dynamics -- at the level of firing rates -- are reliable, but primarily track network input~\cite{vanVreeswijk:1996us,vanVreeswijk:1998uz,Rosenbaum2014,pyle2016highly,Renart2010}.  Biological neural networks generate reliable, intricate responses to simple sensory inputs, for example to produce motor output~\cite{Churchland2012,Shenoy2013}.  
This raises the  question of how neural circuits  reliably produce intricate  firing rate dynamics for dynamical computations.

In this letter, we show that the limited dynamical complexity of firing rates in asynchronous-irregular spiking networks is overcome by incorporating the widely reported dependence of connection probability on distance~\cite{Lund2003,Levy:2012dy,Ercsey-Ravasz2013,knoblauch2014brain}.  Spiking networks with a spatial topology can undergo symmetry-breaking Turing-Hopf bifurcations~\cite{ricard2009turing,roxin2005role} to generate intricate spatiotemporal dynamics that can be trained to perform computations.

\paragraph{Results}
Following previous work~\cite{Rosenbaum2014}, we consider a recurrent neural network with $4\times 10^4$ excitatory ($\ee$) and $10^4$ inhibitory ($\ii$) model neurons \newstuff{arranged uniformly on a square-shaped domain, $\Gamma=[0,1]\times[0,1]$, with periodic boundaries, \emph{i.e.} a torus.} 
The synaptic input current to neuron $j$ in population $a=\ee,\ii$ is given by 
$$
I^a_j(t)=\sum_{k=1}^{N_\ee}\hspace{-.02in} J_{jk}^{a\ee} \sum_n \delta(t-t_{nk}^\ee)+\sum_{k=1}^{N_\ii}\hspace{-.02in} J_{jk}^{a\ii}\sum_n \delta(t-t_{nk}^\ii)+F_j^a(t)
$$
where $t_{nk}^b$ is the $n$th spike of neuron $k$ in population $b=\ee,\ii$.  Spikes are determined by a leaky integrate-and-fire dynamic~\footnote{Membrane potential of neuron $j$ in population $a=\ee,\ii$ obeys $V'=-(V-E_L)/\tau_m+I_j^a$ where $E_L=-70$~mV, $\tau_m=20$~ms and each time $V$ exceeds $V_{th}=-50$~mV, a spike is recorded and $V$ is reset to $V_{re}=-75$~mV.  In all simulations, average connection probabilities were $\overline p_\eeee=\overline p_\iiee=0.0125$, $\overline p_\eeii=\overline p_\iiii=0.05$, so that each neuron received 500 excitatory and 500 inhibitory inputs on average and connection weights were $j_\eeee=0.1$, $j_\iiee=0.2$ and  $j_\eeii=j_\iiii=-0.25$~mV.}.  To model the widely observed distance-dependence of  connection probability~\cite{Lund2003,Levy:2012dy,Ercsey-Ravasz2013,knoblauch2014brain}, the synaptic weight from a neuron at coordinates ${\bf y}\in \Gamma$ in population $b$ to a neuron at  $\bf x\in\Gamma$ in population $a$ is chosen randomly according to
$$
J_{jk}^{ab}=\begin{cases}j_{ab}& \textrm{with prob.~}p_{ab}({\bf x}-{\bf y})\\ 0 & \textrm{otherwise}\end{cases},
$$
where $p_{ab}({\bf u})=\overline p_{ab}G(\vvec u;\sigma_b)$ and $G(\vvec u;\sigma_b)$ is a two-dimensional wrapped Gaussian with width $\sigma_b$~\cite{Rosenbaum2014}.  

\begin{figure}
 \centering{
 \includegraphics[width=85mm]{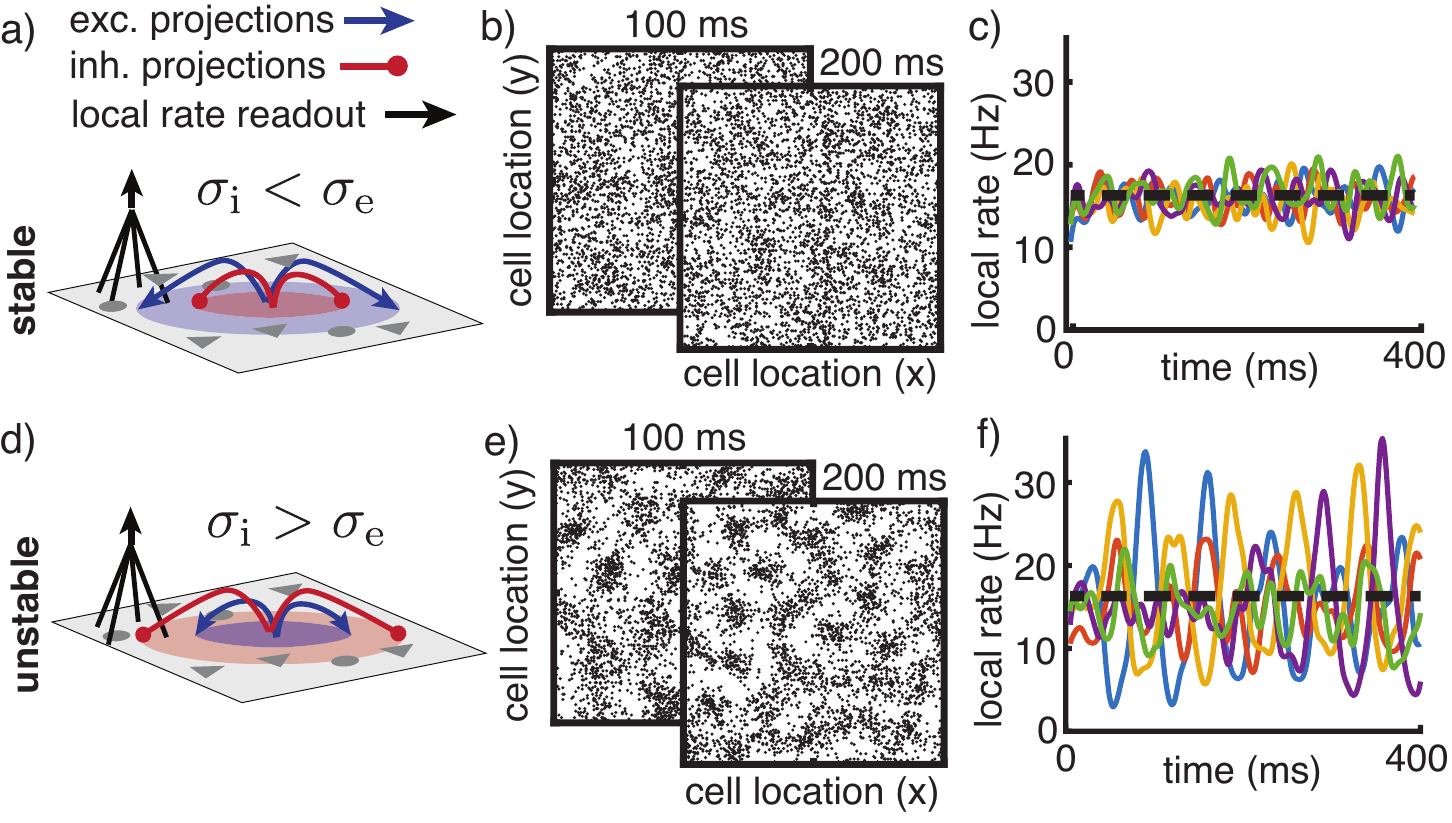}
 }
 \caption{{\bf Intrinsic dynamics in a spatially extended spiking network.} {\bf a)} Schematic of spatially extended spiking network model.  Excitatory and inhibitory neurons arranged on a square project randomly to one another.  Lateral excitatory projections (blue) are are longer on average than lateral inhibitory projections (red; $\sigma_\ee=0.1$, $\sigma_\ii=0.05$, $F^\ee(t)=3$~V/s, $F^\ii(t)=2.3$~V/s).  Each local rate readout (black) is computed by averaging the activity of all excitatory neurons within a square patch then low-pass filtering with a Gaussian kernel ($\sigma=5$~ms). {\bf b)} Raster plot snapshots over 5~ms time windows starting at  $t=100$ and 200~ms. 
 {\bf c)} Five randomly chosen local rate readouts.  Dashed black line shows numerically computed fixed point  rate. 
 {\bf d-f)} Same as a-c, but inhibitory projections are broader than excitatory ($\sigma_\ee=0.05$, $\sigma_\ii=0.1$). 
See Supplemental Animation for animated raster plots.
 }
 \label{F:Rasters}
 \end{figure}

We first simulated a network in which external inputs were constant across space and time 
and inhibitory projections were more localized than excitatory projections (Fig.~\ref{F:Rasters}a).  Even though the model is deterministic, spiking activity was irregular and asynchronous with no coherent spatial patterning (Fig.~\ref{F:Rasters}b, Supplementary Figures 1,2 and Supplementary Animation).  This biologically realistic spike-timing variability is driven by chaos-like dynamics~\cite{vanVreeswijk:1996us,vanVreeswijk:1998uz,Lajoie2013,Rosenbaum2014,Monteforte2012}.

Despite the complexity of spike-timing dynamics, firing rates are amenable to mean-field analysis under a diffusion approximation~\cite{Rosenbaum2014}.  The mean input, $\vec \mu({\bf x})=[\mu_\ee({\bf x})\; \mu_\ii({\bf x})]^T$, to e and i neurons near ${\bf x}\in\Gamma$ is
$$
\vec \mu({\bf x})=
\iint_\Gamma W({\bf u})\vec r({\bf x}-{\bf u})d{\bf u}+\vec F({\bf x})
$$
where $\vec r({\bf x})=[r_\ee({\bf x})\; r_\ii({\bf x})]^T$ is the average firing rate and $\vec F({\bf x})$ the feedforward input to neurons near ${\bf x}\in\Gamma$.  The matrix kernel $W({\bf u})$ captures synaptic divergence and similarly for the input variance, $v({\bf x})=\iint  U({\bf u})\vec r({\bf x}-{\bf u})d{\bf u}$~\footnote{$W=[w_\eeee({\bf u}) \; w_\eeii({\bf u});\; w_\iiee({\bf u})\; w_\iiii({\bf u})]$ is a $2\times 2$ matrix function where $w_{ab}({\bf u})=j_{ab}p_{ab}({\bf u})N_b$ and similarly for $U({\bf u})$, except that $u_{ab}=j_{ab}w_{ab}$.}.  The mapping from input statistics to  rates, $\vec r=\phi(\vec \mu,\vec v)$, is computed using a Fokker-Planck formalism so fixed point rates can be computed numerically~\cite{Amit:1997uj,Richardson:2007ct,richardson2009dynamics,Frontiers16}.  When $\vec F({\bf x})=\vec F$ is spatially uniform, so are fixed point rates~\cite{Rosenbaum2014}.

 To estimate local   rates from simulations, we partitioned the network into 100  squares, then averaged and low-pass filtered the spike trains of the 400 excitatory neurons in each square (Fig.~\ref{F:Rasters}a).  This yields a readout of the local instantaneous firing rates and also models synaptic output that the network would send to downstream neural populations.  These local rate readouts closely matched the fixed point firing rates computed numerically from the diffusion approximation (Fig.~\ref{F:Rasters}c) and fluctuations in the rates were consistent with asynchronous, Poisson-like spike timing variability~\footnote{See Supplementary Figure 1 for a comparison with Poisson spike trains.}.

Broad lateral inhibition is known to induce spatial pattern formation~\cite{ermentrout1998neural,coombes:2005,bressloff:2012,Sadeh2014,Sadeh2015,Rosenbaum2014}.  We next modified the network so that inhibitory projections were broader than excitatory projections (Fig.~\ref{F:Rasters}d).  This  produced a dramatic change in the spiking activity, with spatially uniform activity giving way to intricate, asymmetric spatiotemporal activity patterns  (Fig.~\ref{F:Rasters}e and Supplementary Animation), despite the spatial symmetry of connection probabilities in the network.  These spatiotemporal patterns were reflected in the local rate readouts by irregular high-amplitude fluctuations (Fig.~\ref{F:Rasters}f).   
\newstuff{Despite their differences, both networks produced asynchronous, irregular spike trains with an approximate balance between excitation and inhibition~\footnote{See Supplementary Figure~2.}.}

The network with broad inhibition and the network with local inhibition share the same spatially uniform fixed point  under the diffusion approximation, but rates strongly deviated from this fixed point when inhibition was broader. 
We therefore conjectured that the fixed point was stable for the simulation with local inhibition and unstable when inhibition was broader. 

\begin{figure*}
 \centering{
 \includegraphics[width=160mm]{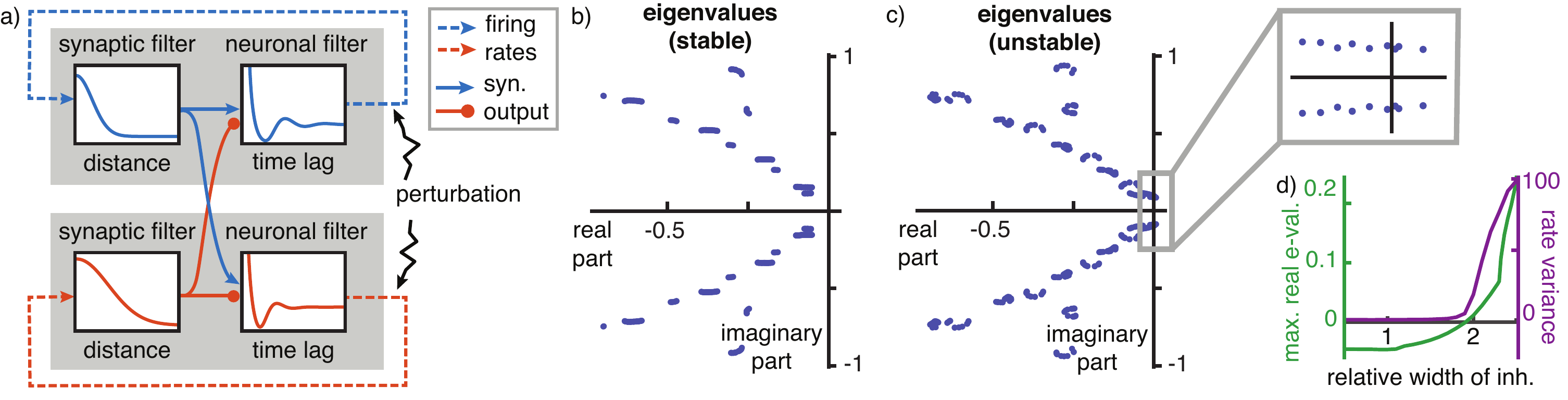}
 }
 \caption{ {\bf Stability of spatially extended spiking networks.} {\bf a)} Schematic of stability analysis. A perturbation applied to the firing rates (dashed lines, blue for excitatory and red for inhibitory) is filtered spatially by synaptic divergence (left boxes, showing connection probability as a function of distance) to determine perturbations of synaptic currents (solid lines), which are filtered temporally by neurons (right boxes, showing linear response kernels). 
{\bf b)}  When inhibition is more local than excitation (as in Fig.~\ref{F:Rasters}a-c), all eigenvalues have negative real parts.  {\bf c)} Same as (b), except for broad inhibition (as in Fig.~\ref{F:Rasters}d-f). Some eigenvalues have positive real part (insert).  {\bf d)} Maximum real part of the eigenvalues (green) and the average temporal variance of the firing rate readouts (purple; units Hz$^2$) as a function of the relative width of inhibitory projections ($\sigma_\ii/\sigma_\ee$).
 }
 \label{F:Stability}
 \end{figure*}

\newstuff{Spatially extended  neural networks are often described using {integro-differential} equations of the form 
\begin{equation}\label{E:WC}
\tau{\partial{\vec r}}/{\partial t}=-\vec r+\phi(\vec \mu,\vec v)
\end{equation}
or of a similar form. 
As in previous work~\cite{Rosenbaum2014}, this approach predicted stability of uniform firing rates for the network with broad inhibition, despite the patterns observed in simulations~\footnote{See Supplementary Note S.1 and Supplementary Figure 4 for a review of the stability analysis of the neural field model.}.  
We conjectured that firing rate dynamics observed when inhibition was broad arose in part from a resonance in neurons' membrane and spiking dynamics~\cite{Lindner2004} that is not captured by Eq.~\eqref{E:WC}.
To account for this resonance, we generalized the stability analysis from recent work~\cite{Ledoux:2011cv} to spatial networks.  
Linear response theory 
gives an integral equation for the dynamics of a perturbation from the fixed point~\footnote{See Supplementary Note S.2 for a derivation.},
\begin{equation}\label{E:op}
\begin{aligned}
\vec r({\bf x},t)=
&\iint_\Gamma \int_0^\infty A(\tau)W({\bf u})\vec r({\bf x}-{\bf u},t-\tau)d\tau d{\bf u}  \\
&+\iint_\Gamma \int_0^\infty B(\tau)U({\bf u})\vec r({\bf x}-{\bf u},t-\tau)d\tau d{\bf u}.  
\end{aligned}
\end{equation}
as illustrated in Fig.~\ref{F:Stability}a.
The matrix kernels, $A(\tau)$ and $B(\tau)$, quantify
excitatory and inhibitory neurons' linear response to perturbations in their input mean and variance~\footnote{$A(\tau)=[A_\ee(\tau) \; 0;\; 0\; A_\ii(\tau)]$ where $A_a(\tau)$ is the excitatory neurons' linear response to a perturbation in mean input, and similarly for  $B(\tau)$.}. 
Eq.~\eqref{E:op}  can capture an arbitrary linear dependence of firing rates on recent history, which is generally not possible  in a finite system of integro-differential equations like Eq.~\eqref{E:WC}.  
Transitioning to the temporal Laplace and spatial Fourier domains in Eq.~\eqref{E:op} gives
\begin{equation}\label{E:EvansFT}
\det\left[\widehat A(\lambda)\widetilde W({\bf n})+\widehat B(\lambda)\widetilde U({\bf n})-Id\right]=0
\end{equation}
where $Id$ is the $2\times 2$ identity matrix, $\widetilde W({\bf n})$ and $\widetilde U({\bf n})$ are Fourier coefficients of $W({\bf u})$ and $U({\bf u}$), and  $\widehat A(\lambda)$ and $\widehat B(\lambda)$ are matrices of susceptibility functions~\footnote{$\widetilde W=[\widetilde w_\eeee\; \widetilde w_\eeii;\; \widetilde w_\iiee \; \widetilde w_\iiii]$ where $\widetilde w_{ab}({\bf n})=j_{ab}N_b\overline p_{ab} \exp(-2\pi^2 \|{\bf n}\|^2\sigma_b^2)$, similarly for $\widetilde U$ and $\widetilde u_{ab}({\bf n})=j_{ab}\widetilde w_{ab}({\bf n})$.  Also, $\widehat A(\lambda)=[\widehat A_\ee(\lambda)\; 0; \; 0 \;\widehat A_\ee(\lambda)]$ where $\widehat A_x(\lambda)$ is the susceptibility of excitatory neurons and similarly for $\widehat B(\lambda)$.}, 
which can be computed under the diffusion approximation using a Fokker-Planck formalism~\cite{risken1989fokker,Richardson:2007ct,richardson2009dynamics,Frontiers16}. 
Solutions, $\lambda$, to Eq.~\eqref{E:EvansFT} are eigenvalues of the rate dynamics and the associated Fourier modes, ${\bf n}$, are eigenmodes.}

Computation of the eigenvalues confirms that the uniform fixed point rates are stable for the network with local inhibition (Fig.~\ref{F:Stability}b), but unstable for the network with broader inhibition (Fig.~\ref{F:Stability}c).  The eigenvalues with positive real part have non-zero imaginary part (Fig.~\ref{F:Stability}c) and are associated with spatially non-uniform eigenmodes (${\bf n}\ne 0$), implying a Turing-Hopf bifurcation that produces spatially coherent, time-varying patterns~\cite{roxin2005role,ricard2009turing}.  
Varying the width of inhibition shows that eigenvalues with positive real part emerge once inhibition is about twice as broad as excitation, coinciding with the emergence of high-amplitude firing rate variability in simulations (Fig.~\ref{F:Stability}d, compare green and purple).  \newstuff{Stability can also be modulated by the strength of external input to inhibitory neurons~\footnote{See Supplementary Figure~3.}, showing that the dynamical state of the network can be controlled by input.}

\begin{figure}
 \centering{
 \includegraphics[width=83mm]{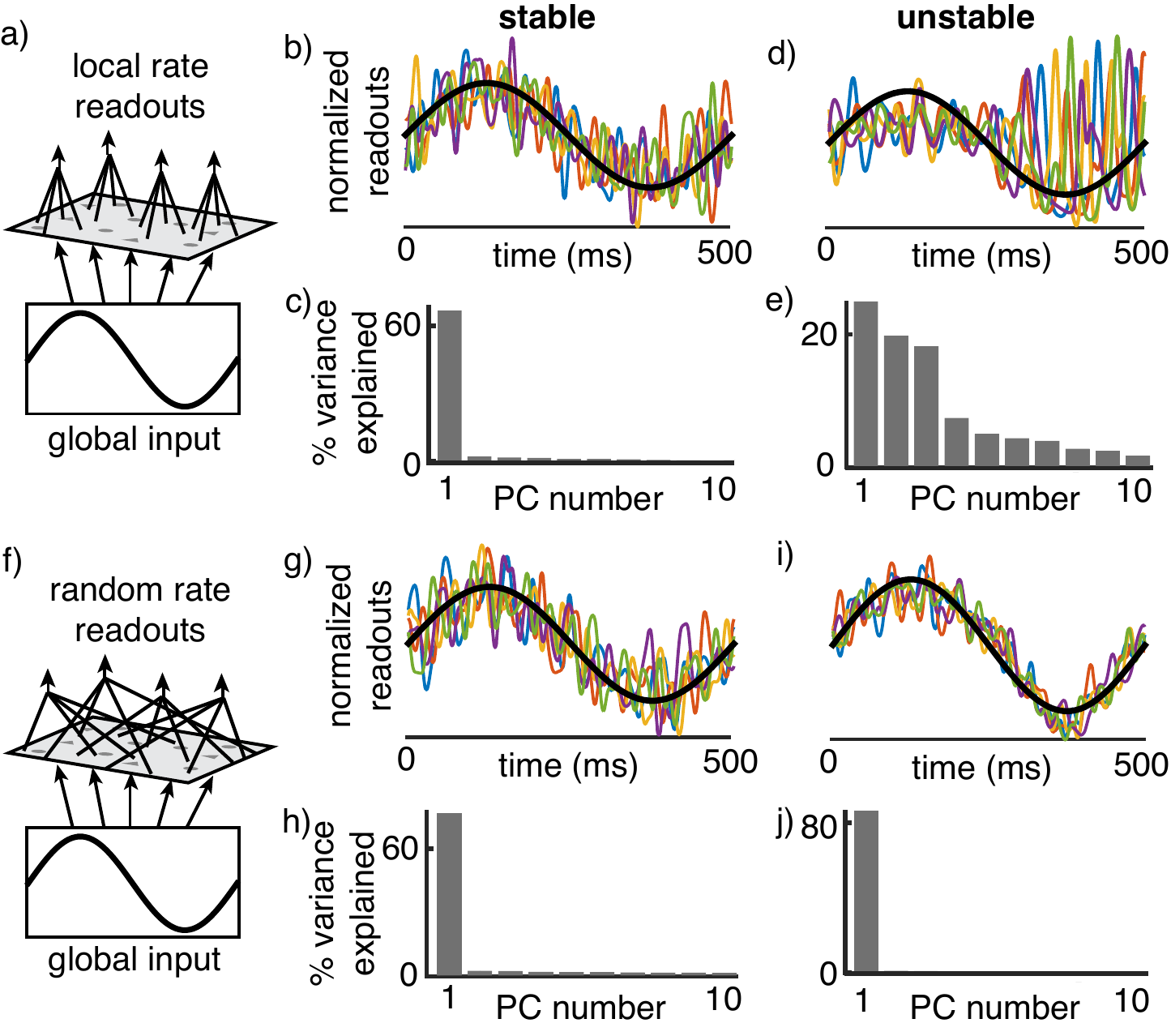}
 }
 \caption{ {\bf Firing rate response to global input.} {\bf a)}  A spatially uniform sinusoidal input was provided to all neurons in the network ($F^\ee(t)=3+1.5\sin(2\pi t)$, $F^\ii(t)=2.3+1.5\sin(2\pi t)$). {\bf b)} Five randomly chosen firing rate readouts (colored curves) track the input (black curve) with Poisson-like variability.  All curves normalized by subtracting their mean and dividing by the standard deviation.  {\bf c)} Percent variance in the 100 firing rate readouts explained by the first ten principal component projections.  {\bf d,e)} Same as b and c, but for the unstable network.  {\bf f-j)} Same as a-e except that firing rates are read out randomly and globally from the network.
 }
 \label{F:GlobalInputBoth}
 \end{figure}

So far, we have considered purely spontaneous firing rate dynamics.
We next added a time-varying external input shared by all neurons in the network (Fig.~\ref{F:GlobalInputBoth}a).  
For the stable network, local firing rates approximately tracked the shared input  with the addition of irregular fluctuations (Fig.~\ref{F:GlobalInputBoth}b), consistent with Poisson-like spike-timing variability~\footnote{See Supplementary Figure 1 for a comparison with Poisson spike trains.}.  
Applying principal component (PC) analysis to the local firing rates revealed that the majority of firing rate variability is captured by the first PC projection (Fig.~\ref{F:GlobalInputBoth}c), representing the variability inherited from the one-dimensional external input.  The remaining variability was spread among higher PC projections, representing unstructured, Poisson-like spike-timing variability.

The unstable  network exhibited a starkly different response to the external input.  While local rates were affected by the external input, they did not reliably track it (Fig.~\ref{F:GlobalInputBoth}d).  The input evoked a high-dimensional response, with variability distributed across several PC projections (Fig.~\ref{F:GlobalInputBoth}e).  These results show that the unstable  network generates high-dimensional firing rate dynamics in response to a  one-dimensional input, while the stable network simply tracks the input with Poisson-like spike timing variability.

Interestingly, while \emph{local} rate readouts of the unstable network did not track the input, random \emph{global} readouts from the same network do track input.  Specifically, we computed firing rate readouts generated from 400 excitatory neurons randomly selected from the entire network (Fig.~\ref{F:GlobalInputBoth}f), instead of locally.   These random readouts from both the stable and unstable networks reliably tracked external input (Fig.~\ref{F:GlobalInputBoth}g-j).  This finding can be understood by noting that the random readouts estimate the network-averaged  rates.  Eq.~\eqref{E:EvansFT} is identical for the stable and unstable networks at the uniform eigenmode, ${\bf n}={\bf 0}$, so the networks have the same eigenvalues at that mode.  Hence, the global average firing rate exhibits similar dynamics in both networks. 
 \begin{figure}
 \centering{
 \includegraphics[width=85mm]{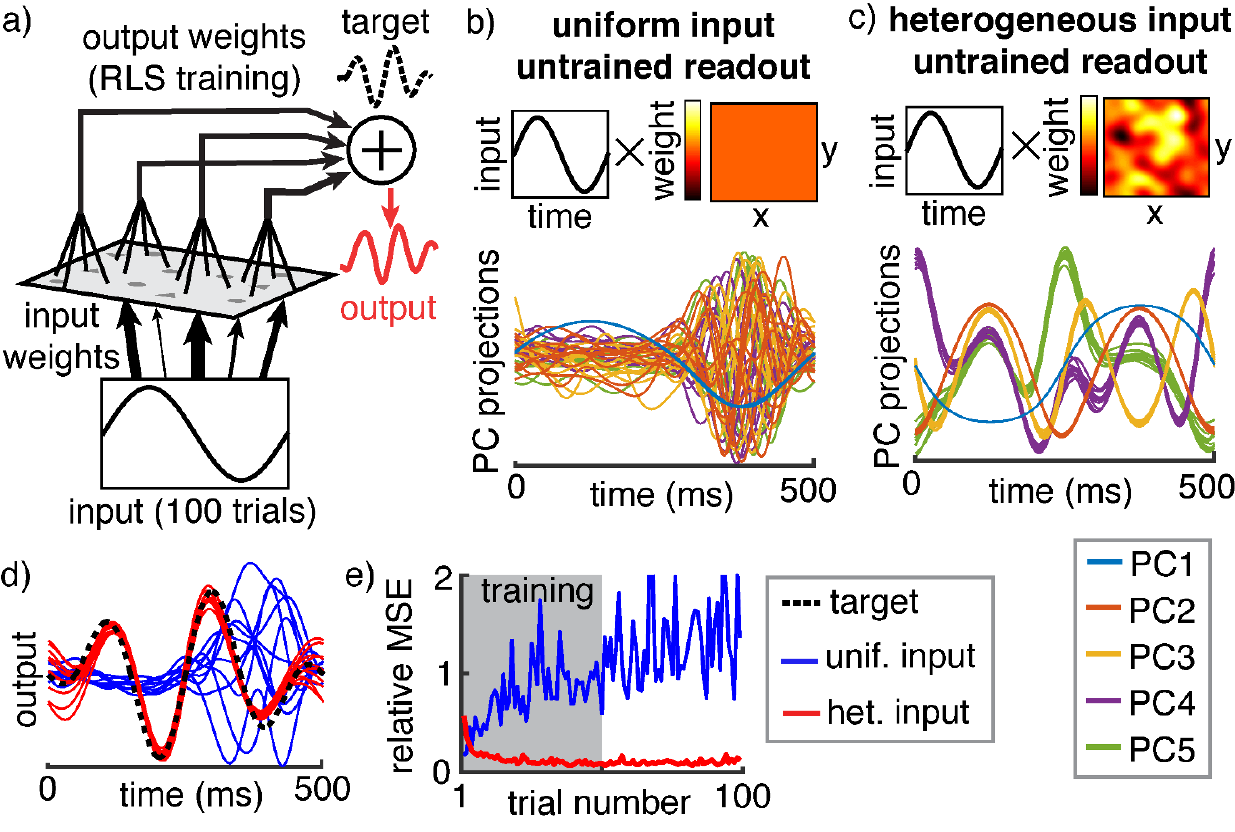}
 }
 \caption{{\bf Reliable computations require heterogeneous input.} {\bf a)} Schematic. Same as Fig.~\ref{F:GlobalInputBoth}a except the input was repeated for 100 consecutive trials and multiplied by fixed, location-dependent weights. Readouts were multiplied by output weights that were trained to produce a target output.    {\bf b)} Overlaid plots of the first five PC projections of untrained readouts (uniform readout weights) over ten randomly selected trials when inputs were spatially uniform (same model as Fig.~\ref{F:GlobalInputBoth}d). {\bf c)} Same as (b) with spatially heterogeneous input weights. {\bf d)} Trained output from last ten trials with uniform (blue) and heterogeneous (red) input weights, compared to target (dashed black).  {\bf e)} Mean-squared error of readouts. 
 }
 \label{F:Reliability}
 \end{figure}

For the firing rate dynamics generated by the unstable network to perform reliable nonlinear computations on inputs, the response of the network should be consistent across repeated presentations of the same input.  
We found that the  transformation of  spatially uniform  input considered in Fig.~\ref{F:GlobalInputBoth}d,e was not reliable: While the first PC projection reliably tracked the input, other components were highly unreliable from trial to trial (Fig.~\ref{F:Reliability}a,b).  We conjectured that this unreliability is due to the spatial symmetry of the network:  Since the activity patterns generated by the unstable network arise through a symmetry-breaking dynamic, each time the input is presented, there are numerous firing rate patterns that are equally likely to be evoked.  As a result, the evoked  response depends on small differences in the network state when the input arrives.

We therefore considered an input that projects to the network with weights that vary across space~\footnote{$F_j^\ee(t)=3+1.5\sin(2\pi t)Q(x,y)$ where  $Q(x,y)$ is a realization of quenched, spatial Gaussian noise with unit variance and $E[Q({\bf x})Q({\bf x}+{\bf u})]= \exp(-|{\bf u}|^2/(2\sigma^2))$ where $\sigma=0.1$.} (Fig.~\ref{F:Reliability}a,c).  This modification had a striking effect on the network response.    Unlike the response to spatially uniform input, the response to spatially heterogeneous input was highly reliable from one presentation of the stimulus to another (Fig.~\ref{F:Reliability}c).

We next asked whether the unstable network could be trained to implement dynamical computations by using the local rate readouts as the ``reservoir'' in a reservoir computing model.  Specifically, the local rate readouts were  linearly combined to produce an output time series.  The weights for the linear combination of readouts were trained using a recursive least-squares algorithm~\cite{Sussillo2009} that iteratively updates weights to mold the output to a target time-series~\footnote{See Supplementary Note S.3 for description of numerical procedure for training weights.}~(Fig.~\ref{F:Reliability}a).  

When this algorithm was applied to the firing rates produced by spatially uniform inputs (from Fig.~\ref{F:Reliability}b), the outputs did not produce the target time series (Fig.~\ref{F:Reliability}d,e, blue curves), due to the unreliability of the network response.  When the same algorithm was applied to the rates produced by spatially heterogeneous inputs (from Fig.~\ref{F:Reliability}c), the outputs closely matched the target (Fig.~\ref{F:Reliability}d,e, red curves).  
Further simulations confirm that the network can produce a variety of target outputs from a variety of inputs~\footnote{See Supplementary Figure 5 for further reservoir computing examples}.

\paragraph{Discussion}

There is an extensive literature on the  analysis of spatially extended neural fields~\cite{ermentrout1998neural,coombes:2005,bressloff:2012} and the dynamics of spiking neuron models~\cite{gerstner2002spiking,izhikevich2007dynamical}, but these  topics are  rarely combined.  Previous studies found spatiotemporal dynamics in spiking networks with synaptic kinetics or  delays~\cite{Lim2013,Sadeh2014,keane2015propagating}. Since the resonance  for a Turing-Hopf bifurcation arises primarily from synaptic dynamics in these models, their stability can be captured by differential neural field equations, similar to Eq.~\eqref{E:WC}.  
The Turing-Hopf bifurcation observed here and in previous work~\cite{Rosenbaum2014} arises from the resonance of spiking neurons, which is not captured by  Eq.~\eqref{E:WC}.  We showed that spatial dynamics arising from the resonance of individual neurons are 
rendered mathematically tractable by extending  linear response techniques developed for homogeneous networks~\cite{Ledoux:2011cv}.  This approach is applicable to the growing class of neuron models for which the linear response function can be computed~\cite{Richardson:2007ct,richardson2009dynamics,Frontiers16}.  

Only a few studies have implemented reservoir computing with spiking  networks.  
Maass et al.~\cite{maass2002real} used a spatially extended spiking network for reservoir computing, but did not explain the role of the spatial topology, which we have clarified.  
More recent studies~\cite{Hennequin2014,Abbott2016} showed that precisely tuning a sub-network of slow synapses offline can produce intricate rate dynamics in spiking networks.   One of those studies~\cite{Abbott2016}, implemented reservoir computing with a spiking network. In the other study~\cite{Hennequin2014} this was only done for a rate network version of the model.  It remains to be shown how this precise tuning of synapses could be achieved biologically, but inhibitory plasticity is one possibility~\cite{Vogels2011}.

Ostojic~\cite{Ostojic2014} showed that spiking networks can produce high-dimensional rate dynamics when synapses are strong, analogous to rate networks with similar structure~\cite{Sompolinsky1988}, but the reliability of these dynamics and their computational capabilities were not explored.  The combination of strong coupling with spatial network topology is a promising direction for future study.

Distance-dependent connectivity is ubiquitous in the brain~\cite{Lund2003,Levy:2012dy,Ercsey-Ravasz2013,knoblauch2014brain}.  We showed that this spatial topology imparts spiking neural networks with the ability to perform dynamical computations (Fig.~\ref{F:Reliability}) while maintaining the ability to accurately track network input (Fig.~\ref{F:GlobalInputBoth}i,j).  Hence, spatial network architecture provides a critical link between biological realism and computational capability in recurrent neural network models.  

\newstuff{
Spatially extended  networks are often modeled with integro-differential equations that do not capture the history-dependence of rate dynamics.  We showed that this shortcoming is overcome using linear response theory to replace the integro-differential equation, \eqref{E:WC}, with an integral equation, \eqref{E:op}.  This approach has applications in any stochastic system with spatially and temporally nonlocal interactions such as models of social networks, population dynamics and epidemiology.}

\begin{acknowledgments}
We thank Ashok Litwin-Kumar, Bard Ermentrout and Brent Doiron  for helpful conversations.  This work was supported by NSF grant DMS-1517828.
\end{acknowledgments}

%

\end{document}